\documentclass[Royal,sagev,times]{sagej}

\usepackage{moreverb,url}
\usepackage{graphicx,amsmath,color, psfrag,amssymb}

\newcommand\BibTeX{{\rmfamily B\kern-.05em \textsc{i\kern-.025em b}\kern-.08em
T\kern-.1667em\lower.7ex\hbox{E}\kern-.125emX}}

\def\volumeyear{2013}

\usepackage[utf8]{inputenc}
\begin{document}

\runninghead{Approximate Bayesian inference for mixture cure models}

\title{Approximate Bayesian inference for mixture cure models}
\author{E. L\'azaro\affilnum{1}, C. Armero\affilnum{1},  V. G\'omez-Rubio\affilnum{2}.}
\affiliation{\affilnum{1}Department of Statistics and Operational Research,  Universitat de Val\`encia,  Doctor Moliner, 50, 46100-Burjassot,  Spain\\
\affilnum{2} Department of Mathematics, Universidad de Castilla-La Mancha Spain.}
\corrauth{Elena L\'azaro\\Department of Statistics and Operational Research,\\  Universitat de Val\`encia,\\  Doctor Moliner, 50, \\ 46100-Burjassot,  Spain}

\email{elena.lazaro@uv.es}

\begin{abstract}

Cure models in survival analysis deal with   populations in which a part of the individuals   cannot experience the event of interest.
Mixture cure models  consider  the target population  as a mixture of susceptible and  non-susceptible individuals. The statistical analysis of these models focuses on  examining the probability of cure (incidence model) and  inferring on the time-to-event in the susceptible subpopulation (latency model).

Bayesian inference on mixture cure models has typically relied upon Markov
chain Monte Carlo (MCMC) methods.
The integrated nested Laplace approximation (INLA) is a recent and attractive approach for doing Bayesian inference.  INLA in its natural definition cannot fit mixture models but recent research has new proposals that combine INLA and  MCMC methods  to extend  its applicability to them~\citep{Bivand2014, GomezRubio2017, GomezRubio2018}.

This paper focuses on the implementation of INLA in mixture cure models. A  general mixture cure survival model  with covariate information for the latency and the incidence model within  a general scenario with  censored and non-censored information is discussed.  The fact that   non-censored individuals  undoubtedly belong   to the uncured population is a valuable information that was  incorporated in the inferential process.

\end{abstract}


\maketitle

\section{Introduction}
Survival analysis is an area of statistics dedicated to researching time-to-event data.
This is  one of the oldest areas of statistics, which dates back to the 1600s with the construction  of  life tables.
 The study of time-to-event data seems simple and traditional  because its main focus is centered on   non-negative random variables. But this is very far from being the case.    The fact that survival times are always positive keeps it away from the normal distribution framework, censoring and truncation schemes  produce non-traditional likelihood issues, and   the  special elements that generate the dynamic nature of events occurring in time make survival analysis an interesting and exciting area of research and application, mainly in the biomedical field.

 Cure models in survival analysis deal with target populations in which a part of the individuals   cannot experience the event of interest. This type of models have largely been developed as a consequence of the  discovery and development of new treatments against cancer.
 The rationale of considering a cure subpopulation comes from the idea that a successful treatment   removes totally the original tumor and the individual cannot    experience any recurrence of the disease. These models allow   to estimate the probability of cure, a key  and valuable outcome in cancer research. This is not the case for  the traditional survival models which consider that all the individuals in the population are at risk.
As stated by  \cite{Lambert2007},  it is important to bear in mind that  cure is considered from  a statistical, population point of view and not from an individual perspective.

Mixture cure models are the most popular cure models. They consider that the target population  is a mixture of susceptible and  non-susceptible individuals. The main interest focuses on  the so called  incidence model,  that accounts for the probability of cure, and the latency model  for the time-to-event in the susceptible subpopulation. A mixture model such as this is very attractive, easy to interpret, and allows to account for model complexity (frailties, time-dependent covariates, etc) in both incidence and latency terms \citep{Peng2014}. Some studies  in cancer research with this type of models are \cite{Sposto2002} who discussed data from trials in paediatric cancer conducted
by the Children’s Cancer Group, \citep{Rondeau2013} who studied recurrences for breast cancer and readmissions for colorectal cancer, and \citep{Hurtado2016}  who centered on melanoma cancer. A very interesting review of these models up to date is \citep{Peng2014}. Cured models also appear  in other areas of research. This is the case of split population
models in economics \citep{Schmidt1989} and limited-failure population life models in reliability   \citep{Meeker1987}.

Bayesian inference always  expresses   uncertainty
in terms of probability distributions \citep{Loredo1989, Loredo1992} and uses Bayes' theorem as often as necessary  in a sequential way to update all relevant information. Bayesian methodology is especially attractive for survival analysis due to its natural treatment of censoring and truncation  schemes as well as the probabilistic quantification of relevant survival outcomes, such as survival probabilities, that they do not need to resort to  asymptotic tools \citep{Ibrahim2005}.

Computation in Bayesian inference is a key issue that allows  the approximate implementation of non-analytical posterior distributions. The integrated nested Laplace approximation (INLA)~\citep{Rue2009} is a recent methodology for doing approximate Bayesian inference in the framework of latent Gaussian models (LGM)~\citep{RueHeld2005}. These   models are a special class of Bayesian additive models  that   cover a wide range of studies and applications~\citep{Rue2017}, and survival models in particular \citep{Martino2011}. INLA, in comparison to Markov chain Monte Carlo (MCMC) methods, provides accurate and fast approximations to the relevant posterior marginal distributions.

INLA is  very attractive and has very good properties but it also has   some limitations. In particular, INLA cannot fit mixture models \citep{Marin2005} in a natural way  because they are generally defined in terms of a combination of different distributions \citep{GomezRubio2017}.
But in science, every constraint or difficulty becomes an opportunity for learning. On this matter, \citep{Bivand2014} and \citep{GomezRubio2017} propose the combination of INLA within  MCMC for mixture models, in particular Gibbs sampling, and fit with INLA   the relevant posterior conditional distributions. \citep{GomezRubio2018} extends these proposals and introduce Modal Gibbs sampling to accelerate the inferential process.

This paper focuses on the implementation of INLA in mixture cure models. A  general mixture cure survival model  with covariate information for the latency and the incidence model within  a general scenario with  censored and non-censored information is discussed.  The fact that   non-censored individuals  undoubtedly belong   to the uncured population is a valuable information that was incorporated in the inferential process.

The organization of this paper is as follows. Section 2 presents the main elements of mixture cure models and the two most popular mixture cure models,    the Cox proportional hazards and the accelerated failure times   models.  Section 3 introduces the integrated nested Laplace
approximation within the general  framework of Bayesian inference. Section 4 is the core of the paper and contains our INLA proposal for estimating mixture cure models. Section 5 applies our   proposal to  the statistical analysis of     two benchmark data sets in the framework of clinical trials and bone marrow transplants, and discusses and compares the subsequent results with those from a MCMC implementation. The paper ends with some conclusions.

\section{Mixture cure models}
\label{s:mcm}
Let $T^*$ be a continuous and non-negative random variable that describes the time-to-event of an individual  in some target population. Let $Z$ be a cure random variable defined as $Z=0$ if that individual is susceptible  for experiencing the event of interest, and $Z=1$ if she/he is cured or immune   for that event. Cure and non cure probabilities are   $P(Z=1)=\eta$ and  $P(Z=0)=1-\eta$, respectively. The survival function for individuals in the cured and uncured population, $S_c(t)$ and $S_u(t)$, $t>0$, respectively,  is
\begin{align}
\label{eqn:mix1}
S_u(t) = & \,\,P(T^* > t \mid Z=0)   \nonumber\\
 S_c(t) = & \,\,P(T^* > t \mid Z=1) = 1.
\end{align}
\noindent The general survival function for $T^*$ can be expressed in terms of a mixture of both cured and uncured populations in the form
\begin{equation}
\label{eqn:mix2}
S(t) =   \,\,P(T^* > t) =  \eta + (1-\eta) \,S_u(t).
\end{equation}
\noindent It is important to point out that $S_u(t)$ is a proper survival function but  $S(t)$ is not. It    goes to $\eta$ and not to zero when $t$ goes to infinity. Cure fraction $\eta$    is also known as the incidence model and  time-to-event $T^*_u$   in the uncured population as the latency model~\citep{Peng2014}.

\subsection{Covariates in the incidence model}

The effect  of a  baseline covariate vector $\boldsymbol{x}_1$ on the cure proportion is typically modeled by means of a logistic link
function,  $\mbox{logit}[\eta(\boldsymbol{\beta}_1 )] = \boldsymbol{\beta}_1^{\prime} \boldsymbol{x}_1$,   also expressed as
\begin{equation}
\label{eqn:mix4}
\eta(\boldsymbol{\beta}_1 ) = \dfrac{\mbox{exp} \{ \boldsymbol{\beta}_1^{\prime}\boldsymbol{x}_1\}}{1+\mbox{exp}\{ \boldsymbol{\beta}_1^{\prime}\boldsymbol{x}_1\}},
\end{equation}
\noindent where $\boldsymbol{\beta}_1$ is the vector of regression coefficients associated to    $\boldsymbol{x}_1$.  Note that other link functions can be used to connect the cure fraction  with the vector of covariates $\boldsymbol{x}_1$ such as   the probit link or the complementary log-log link (see~\cite{Robinson2014} for more details).

\subsection{Covariates in the latency model}

The most common   regression models in survival analysis are the  Cox proportional hazards model~\citep{Cox1972} and the accelerated failure time models. We will introduce them below.\vspace*{0.2cm}\\
 \textsf{Cox proportional hazards model, CPH}. It is usually formulated in terms of the hazard function for the time-to-event  $T_u^*$, or instantaneous rate of occurrence of the event,  as
 \begin{align}
\label{eqn:mix8}
h_{u}(t\mid h_{u0}, \boldsymbol{\beta}_2 )   & = \,\lim_{\Delta t \to \infty} \frac{P(t \leq T_u^{*}<t+\Delta t \mid T \geq t)}{\Delta t} \nonumber \\
                                            & = \,h_{u0}(t) \,\mbox{exp}\{\boldsymbol{\beta}_2^{\prime}\boldsymbol{x}_2\},
\end{align}
\noindent where $h_{u0}(t)$ is the baseline hazard function that  determines the shape of the hazard function.
   Model (\ref{eqn:mix8}) can also be presented in terms of the survival function of $T^*_u$ as
\begin{equation}
\label{eqn:mix9}
S_{u}(t\mid  S_{u0}, \boldsymbol{\beta}_2 ) = [S_{u0}(t)]^{\mbox{\footnotesize{exp}}\{\boldsymbol{\beta}_2^{\prime}\boldsymbol{x}_2\}},
\end{equation}
\noindent where $S_{u0}(t)=\mbox{exp}\{-\int_0^t h_{u0}(s)\mbox{d}s\}$ represents the survival baseline function.

Fully Bayesian methods specify a model for $h_{u0}(t)$ which   may be of  parametric or non-parametric  nature. Exponential, Weibull and Gompertz hazard functions are common parametric   proposals in the empirical literature.  Mixture of piecewise constant functions or   B-splines basis functions are the usual counterpart  in non-parametric selections. They provide a great flexibility to the modeling by allowing different patterns and multimodalities but some care is needed when working with them to avoid overfitting. To this effect, the elicitation of prior distributions is a relevant issue in the Bayesian approach to regularization~\cite{Lazaro2018}.\vspace*{0.25cm}\\
 \textsf{Accelerated failure time models, AFT.}   These models try to adapt the philosophy of linear models to the survival framework. The survival variable  $T_u^*$   is now expressed in the logarithmic scale to extend  the modeling to the real line. It is modeled as the sum of a linear term for the covariates $\boldsymbol{x}_2$, which usually includes an intercept element, and a random error   $\epsilon$ amplified or reduced by a scale factor $\sigma$ as followss
\begin{equation}
\label{eqn:mix5}
\mbox{log}(T_u^*)= \boldsymbol{x}_2^{\prime} \boldsymbol{\beta_2} + \sigma \epsilon.
\end{equation}
Common distributions for $\epsilon$ are normal, logistic and standard Gumbel. They respectively imply   log-normal, log-logistic and Weibull distributions for $T_u^*$~\citep{Christensen2011}. Weibull AFT  models  are the most popular ones, in which covariates  $\boldsymbol{x}_2$  are commonly
included in the scale parameter as $\lambda(\boldsymbol{\beta}_2 )=\mbox{exp}\{\boldsymbol{\beta}_2^{\prime}\boldsymbol{x}_2\}$, and consequently
\begin{align}
\label{eqn:mix7}
h_u(t \mid \alpha,  \boldsymbol{\beta}_2)&=   \,\alpha \,t^{\alpha-1} \,\mbox{exp}\{\boldsymbol{\beta}^{\prime}\boldsymbol{x}_2\}, \nonumber \\
S_u(t \mid \alpha, \boldsymbol{\beta}_2)&=\mbox{exp}\{-t^{\alpha}\mbox{e}^{\{\boldsymbol{\beta}_2^{\prime}\boldsymbol{x}_2\} }\}.
\end{align}
This modeling strategy  based on introducing covariate information through one of the parameters of the target distribution also applies to  the rest of parametric probability distributions.

%

 \section{Bayesian inference and the integrated nested Laplace approximation}
\label{s:inla}

 Bayesian inference derives the posterior distribution of the quantities of interest according to Bayes' theorem, which combines  the prior distribution  of all unknown quantities and the likelihood function   constructed   from the data.  It is the main  element in Bayesian statistics and   starting point of all relevant inferences. The posterior distribution in complex  models is   non analytical  and for this reason  it   needs to be  computationally approached. To that effect,  MCMC methods are surely the most popular procedures    although they  involve large computational costs and require additional work for checking convergence and accuracy  estimation.

The structure and main elements of the INLA approach for doing Bayesian inference are summarised below. Let us assume a set of $n$ random variables $\boldsymbol T^*\,=\,(T^*_1,\ldots,T^*_n)$   mutually  conditionally independent  given a  latent Gaussian Markov random field (GMRF)~\citep{RueHeld2005} $\boldsymbol \theta$
and a set of likelihood hyperparameters $\boldsymbol \phi_2$. The GMRF $\boldsymbol \theta$ depends on some hyperparameters  $\boldsymbol \phi_1$ and can include  effects of different type (regression coefficients, random effects, seasonal effects, etc).

According to Bayes' theorem, the joint posterior distribution for $(\boldsymbol \theta, \boldsymbol \phi)$, where $\boldsymbol \phi\,=\,(\boldsymbol \phi_{1},\boldsymbol \phi_{2})$, after data $\mathcal{D}=\cup_{i=1}^{n}\, \mathcal{D}_{i}$ have been observed, where $\mathcal{D}_i$ represents the data from individual $i$th,  can be written as
\begin{align}
\label{eqn:inla2a}
\pi(\boldsymbol \theta, \boldsymbol \phi \mid \mathcal{D})\,& \propto\,\mbox{$\prod$}_{i=1}^{n} \,\mathcal{L}_i(\boldsymbol \theta, \boldsymbol \phi \mid \mathcal{D}) \,  \pi(\boldsymbol \theta,  \boldsymbol \phi) \nonumber\noindent \\
                                      &\propto\,\mbox{$\prod$}_{i=1}^{n} \,\mathcal{L}_i(\boldsymbol \theta, \boldsymbol \phi \mid \mathcal{D} )\,\pi(\boldsymbol \theta \mid  \boldsymbol \phi)\,\pi(\boldsymbol \phi),
\end{align}
\noindent where $\mathcal{L}_i(\boldsymbol \theta, \boldsymbol \phi \mid \mathcal{D})$ is the likelihood function of $(\boldsymbol \theta, \boldsymbol \phi)$ for data $\mathcal{D}_i$, and $\pi(\boldsymbol \theta, \boldsymbol \phi)$  represents the prior distribution of $(\boldsymbol \theta, \boldsymbol \phi)$ which   factorizes as the product of a GMRF conditional prior distribution
$\pi(\boldsymbol \theta \mid  \boldsymbol \phi)$ and a  marginal prior distribution $\pi(\boldsymbol \phi)$.

INLA
makes use of  Laplace approximations~\citep{Rue2009} to   obtain approximations $\tilde{\pi}(\boldsymbol \phi \mid \mathcal{D})$ and $\tilde{\pi}(\theta_{\hbox{\scalebox{1.3}{$\cdot$}}} \mid \boldsymbol \phi, \mathcal{D})$ for the posterior distribution  $\pi(\boldsymbol \phi \mid \mathcal{D} )$ and $\pi(\theta_{\hbox{\scalebox{1.3}{$\cdot$}}} \mid \boldsymbol \phi, \mathcal{D} )$, respectively, where
$\theta_{\hbox{\scalebox{1.3}{$\cdot$}}}$ denotes  a generic univariate element in $\boldsymbol \theta$. The marginal posterior distribution for the latent terms $\pi(\theta_{\hbox{\scalebox{1.3}{$\cdot$}}}\mid \mathcal{D} )$  can be obtained as
\begin{align}
\label{eqn:inla3}
&\pi(\theta_{\hbox{\scalebox{1.3}{$\cdot$}}} \mid \mathcal{D} )= \mbox{$\int$}\, \pi(\theta_{\hbox{\scalebox{1.3}{$\cdot$}}}\mid  \boldsymbol \phi, \mathcal{D} )\, \pi(\boldsymbol \phi \mid \mathcal{D} )\,\mbox{d}\boldsymbol \phi,
\end{align}

\noindent and consequently, it can be approximated  by numerical integration  as
\begin{equation}
\label{eqn:inla4}
\tilde{\pi}(\theta_{\hbox{\scalebox{1.3}{$\cdot$}}} \mid \mathcal{D} )  \approx \sum_m  \tilde{\pi}(\theta_{\hbox{\scalebox{1.3}{$\cdot$}}} \mid  \boldsymbol \phi_m, \mathcal{D} )\,\tilde{\pi}(\boldsymbol \phi_m \mid  \mathcal{D} )\, \Delta_m,
\end{equation}

\noindent where $\boldsymbol \phi_m$ are points   in the hyperparametric  space $\Phi$, and $\Delta_{m}$   integration weights. The posterior marginal distribution   $\pi(\phi_{\hbox{\scalebox{1.3}{$\cdot$}}} | \mathcal{D}^{\prime})$  can also be approximated by numerical integration according to the expression
\begin{align}
\label{eqn:inla5}
&\pi(\phi_{\hbox{\scalebox{1.3}{$\cdot$}}} \mid \mathcal{D} )= \mbox{$\int$}\, \pi( \boldsymbol \phi \mid \mathcal{D} )\, \mbox{d}\boldsymbol \phi_{-{\hbox{\scalebox{1.3}{$\cdot$}}}},
\end{align}
\noindent where $\boldsymbol \phi_{-{\hbox{\scalebox{1.5}{$\cdot$}}}}$ represents all elements in $\boldsymbol \phi$ except $\phi_{\hbox{\scalebox{1.5}{$\cdot$}}}$.

 INLA  is implemented in the package \texttt{R-inla} for the R statistical software~\citep{R2014}. This package implements a number of latent effects and allows for an easy model fitting and visualization of the output. A recent review on INLA and the  can be found in~\citep{Rue2017}.


%
\section{INLA to estimate mixture cure models}
In general, standard survival models such as  CPH    and  AFT models can be expressed in terms of GMRF models, and consequently they can be adapted for its INLA implementation\cite{Akerkar2010,Martino2011}. In the case of CPH models, the baseline hazard function is reparameterized in  the exponential scale in order to be included in the CPH element that accounts for regression information. This exponential term also allows   the inclusion of time-varying covariate effects,   nonlinear, structured or non random effects, spatial modelling, etc~\citep{Hennerfeind2006}. They can be expressed by means of a structured geoadditive predictor  whose elements can be modeled in terms of a GMRF model.   AFT models also have this nice relationship and behaviour for INLA implementation.

\subsection{Gibbs sampler for mixture estimation}

Let us consider a general   survival scenario in the framework of    non-informative and independent right censoring and a mixed cure sampling model.
Survival time   is defined as the pair $(T, \delta)$, where $T=\mbox{min}(T^{*}, C)$, $C$ being the censoring time, and $\delta$ an indicator function  defined as $\delta=0$ when the subsequent observation is censored ($T^{*}>C$), and $\delta=1$ when it is not. We assume that the distribution of $T^*$ depends on a conditional GRMF $\boldsymbol \theta$ on hyperparameters $\boldsymbol \phi_1$  and a likelihood hyperparametric vector $\boldsymbol \phi_2$, and consider $\pi(\boldsymbol \theta, \boldsymbol \phi)$ as the prior  distribution for $ (\boldsymbol \theta, \boldsymbol \phi)$ which factorizes as
\begin{equation}
\pi(\boldsymbol \theta, \boldsymbol \phi)=\pi(\boldsymbol \theta \mid \boldsymbol \phi)\,\pi(\boldsymbol \phi).
\end{equation}
Let  $\mathcal{D}_{i} =(t_i, \delta_i)$  represent the survival observed data for individual $i$, $i=1,\ldots,n$, and  $\mathcal{D}=\cup_{i=1}^{n} \mathcal{D}_{i} $. The complete data for individual $i$ is defined as $\mathcal{D}_{com,i}=(t_i, \delta_i,  z_i)=(\mathcal{D}_i, z_i)$, which  includes the value $z_i$ of the subsequent latent variable that classifies this individual as cured or not,  and   $\mathcal{D}_{com}=\cup_{i=1}^{n} \mathcal{D}_{com,i}$.
It should be  noted that an observed  survival time clearly indicates that the subsequent individual belongs to the uncured   population.

The complete data likelihood function  is the product of the complete likelihood function for each individual defined as~\cite{Ibrahim2005}
\begin{equation}
\begin{split}
\mathcal{L}(\boldsymbol{\theta}, & \boldsymbol \phi \mid \mathcal{D}_{com})  =\\
& = \,\mbox{$\prod_{i=1}^{n}$}\,  \mathcal{L}_i(\boldsymbol{\theta}, \boldsymbol \phi \mid \mathcal{D}_{com}) \\
& = \,  \mbox{$\prod_{i=1}^{n}$}\, \eta_i(\boldsymbol \theta, \boldsymbol \phi)^{z_{i}}\,(1-\eta_i(\boldsymbol \theta, \boldsymbol \phi))^{1-z_{i}}\,h_{iu}(t_i\mid \boldsymbol \theta, \boldsymbol \phi)^{\delta_{i}\,(1-z_i)}\,S_{iu}(t_i\mid \boldsymbol \theta, \boldsymbol \phi)^{(1-z_i)}.
\end{split}
\end{equation}
As $\boldsymbol z$ is seldom observed, it is often treated as another parameter in the model and its posterior distribution needs to be computed as well.
The posterior distribution for $(\boldsymbol \theta, \boldsymbol \phi)$ computed from Bayes' theorem would be
\begin{align*}
\pi(\boldsymbol \theta, \boldsymbol \phi \mid \mathcal{D}) \propto \, & \,\mathcal{L}(\boldsymbol{\theta}, \boldsymbol \phi \mid \mathcal{D})\, \pi(\boldsymbol \theta, \boldsymbol \phi)\nonumber\\
\propto \,& \mbox{$\sum$}_{\boldsymbol z \in \mathcal{Z}}\,\,\mathcal{L}(\boldsymbol{\theta}, \boldsymbol \phi \mid \mathcal{D}, \boldsymbol z)\, \pi(\boldsymbol \theta, \boldsymbol \phi),
\end{align*}
where $\mathcal{L}(\boldsymbol{\theta}, \boldsymbol \phi \mid \mathcal{D})$ is the likelihood function of $(\boldsymbol \theta, \boldsymbol \phi )$ for the observed data $\mathcal{D}$, and  $\mathcal{Z}$ denotes the parameter space of the cure indicator values, which is the $n$-dimensional Cartesian product of the binary set $\{0, 1\}$.

The introduction of the latent indicator in the inferential process  and the  Gibbs sampler is the usual procedure to approach Bayesian mixture estimation \citep{Diebolt1994, Marin2005}. We follow this proposal  and consider the inferential process defined by the joint posterior distribution
\begin{align*}
\pi(\boldsymbol \theta, \boldsymbol \phi, \boldsymbol z  \mid \mathcal{D}) \propto \, & \,\mathcal{L}(\boldsymbol{\theta}, \boldsymbol \phi \mid \mathcal{D}_{com})\, \pi(\boldsymbol \theta, \boldsymbol \phi),
\end{align*}
 and a Gibbs sampler based on the full conditional posterior distributions $\pi(\boldsymbol \theta, \boldsymbol \phi  \mid \boldsymbol z, \mathcal{D})$ and $\pi(\boldsymbol z \mid \boldsymbol \theta, \boldsymbol \phi,  \mathcal{D})$.

\subsection{INLA  and modal Gibbs}

Our proposal for fitting  mixture cure models by means of INLA is based on  \cite{GomezRubio2017} and  \cite{GomezRubio2018}, who  use   INLA for estimating the conditional posterior marginals of the model parameters $\pi(\theta_{\hbox{\scalebox{1.3}{$\cdot$}}} \mid \mathcal{D}, z )$ and $\pi(\phi_{\hbox{\scalebox{1.3}{$\cdot$}}} \mid \mathcal{D}, z)$, which  assumes that  the latent vectors which determine the subpopulation to which each individual belongs to are known.  All relevant  marginal posterior distributions, $\pi(\theta_{\hbox{\scalebox{1.3}{$\cdot$}}} \mid  \mathcal{D}_{com})$ and $\pi(\phi_{\hbox{\scalebox{1.3}{$\cdot$}}} \mid  \mathcal{D}_{com})$, can be fitted as usual in the INLA approach.

 The   posterior marginal distribution for each   $\theta_{\hbox{\scalebox{1.3}{$\cdot$}}}$ can be computed as
\begin{align}
\label{eqn:inla}
\pi(\theta_{\hbox{\scalebox{1.3}{$\cdot$}}} \mid \mathcal{D} )  = & \sum_{\boldsymbol{z} \in \mathcal{Z}}  \,\, \pi(\theta_{\hbox{\scalebox{1.3}{$\cdot$}}}, \boldsymbol{z}  \mid  \mathcal{D} )\nonumber\\
     = & \sum_{\boldsymbol{z} \in \mathcal{Z}}  \,\, \pi(\theta_{\hbox{\scalebox{1.3}{$\cdot$}}} \mid \boldsymbol{z},  \mathcal{D} )\,\pi(\boldsymbol{z}  \mid  \mathcal{D} ) \nonumber\\
     = & \sum_{\boldsymbol{z} \in \mathcal{Z}}  \,\,   \pi(\theta_{\hbox{\scalebox{1.3}{$\cdot$}}} \mid  \mathcal{D}_{com})\,\pi(  \boldsymbol{z}  \mid  \mathcal{D}),
\end{align}
\noindent where $\pi(\theta_{\hbox{\scalebox{1.3}{$\cdot$}}} \mid \boldsymbol z, \mathcal{D})$  is fitted by INLA and  $\pi(\boldsymbol{z}  \mid  \mathcal{D} )$ is the marginal  posterior distribution for the latent cure indicator vector  based on the observed data. This latter distribution will be computed using modal Gibbs sampling as proposed by \cite{GomezRubio2018}. The computation of $\pi( \phi_{\hbox{\scalebox{1.3}{$\cdot$}}} \mid \mathcal{D})$ follows a similar  procedure.

Expression (\ref{eqn:inla}) needs some additional discussion so that it can be better   adapted to the cure models framework. Here, we know that each survival observation can be censored or uncensored. In the case of a censored data, we do not know if the subsequent  individual  can or cannot experience the event of interest, hence their belonging to the uncured or  cured subpopulation is unknown and consequently, there will be uncertainty about the value of the corresponding cure indicator variable. Conversely, an uncensored observation   will indicate that the subsequent individual has surely experienced the event of interest, and therefore she/he  belongs to the uncured subpopulation. If we split $\boldsymbol{z}=(\boldsymbol{z}_{unc}, \boldsymbol{z}_{cen})$, where $\boldsymbol{z}_{unc}$ ($\boldsymbol{z}_{cen}$) represents the $n_{unc}$ ($n_{cen}$)-dimensional latent cure indicator corresponding to the uncensored (censored) data, the complete knowledge on the value of the latent indicator of the uncensored data will imply $\boldsymbol{z}_{unc}=\boldsymbol 0$. For this reason,
\[
 \pi(  \boldsymbol{z}  \mid  \mathcal{D} ) = \pi(  \boldsymbol{z}_{unc}, \boldsymbol z_{cen}  \mid  \mathcal{D} ) =
  \begin{cases}
  \pi(  \boldsymbol{z}_{cen}  \mid  \mathcal{D} ) & \text{for } \boldsymbol z_{unc}=\boldsymbol 0, \, \boldsymbol z_{cen} \in \mathcal{Z}_{cen}    \\
   0     & \text{otherwise, }
  \end{cases}
\]
\noindent where now $\mathcal{Z}_{cen}$ is the parameter space of the cure indicator variables for the censored observations,   with lower dimensionality than $\mathcal{Z}$. Hence, expression (\ref{eqn:inla}) can be rewritten as
\begin{equation}
\label{eqn:inla2b}
\pi(\theta_{\hbox{\scalebox{1.3}{$\cdot$}}} \mid \mathcal{D} )\,=\,  \sum_{\boldsymbol{z}_{cen} \in \mathcal{Z}_{cen}}  \pi(\theta_{\hbox{\scalebox{1.3}{$\cdot$}}} \mid  \mathcal{D}_{com})\,\pi(  \boldsymbol{z}_{cen}  \mid  \mathcal{D}).
\end{equation}


The above procedure can be described  in a more structured way via the following algorithm:
\begin{itemize}
  \item[]\textsf{Step 0.}  Assign initial values to the latent cure indicator of the $n_{cen}$ censored observations, $\boldsymbol z_{cen}^{(0)}$, and  consider $\boldsymbol z_{unc}=\boldsymbol 0$ for the uncensored observations. Define $\boldsymbol z^{(0)}=\{ \boldsymbol z_{cen}^{(0)}, \boldsymbol z_{unc} \}$.\vspace*{-0.3cm}\\
  \item[]\textsf{Step 1.}   For $m\,=\,1,2,\ldots $
 \begin{itemize}\vspace*{-0.1cm}
 \item[(a)]Use INLA to approximate $\pi(\theta_{\hbox{\scalebox{1.3}{$\cdot$}}} \mid \boldsymbol{z}^{(m-1)}, \mathcal{D} )$   and $\pi(\phi_{\hbox{\scalebox{1.3}{$\cdot$}}} \mid  \boldsymbol{z}^{(m-1)}, \mathcal{D})$.
      \item[(b)]Compute the subsequent posterior (conditional) modes $ \hat{\theta}_{\hbox{\scalebox{1.3}{$\cdot$}}}^{(m)} $  and $ \hat{\phi}_{\hbox{\scalebox{1.3}{$\cdot$}}}^{(m)}$, respectively, from each of the posterior distributions in (a).
 \item[(b)] Sample $\boldsymbol z_{cen}^{(m)}=(z_{cen,1}^{(m)},  \ldots, z_{cen,n_{cen}}^{(m)})$ from the full conditional distribution for the cure latent variable~\citep{Marin2005, Caia2012}
     \begin{align*}
\label{eqn:uncured}
\pi(Z =0&\mid \mathcal{D}, \boldsymbol{\hat{\theta}}^{(m)}, \boldsymbol{\hat{\phi}}^{(m)}) =\\
 =&\dfrac{(1-\eta_c(\boldsymbol{\hat{\theta}}^{(m)}, \boldsymbol{\hat{\phi}}^{(m)})   \,S_{cu}(t_c\mid \boldsymbol{\hat{\theta}}^{(m)}, \boldsymbol{\hat{\phi}}^{(m)})}{\eta_c(\boldsymbol{\hat{\theta}}^{(m)}, \boldsymbol{\hat{\phi}}^{(m)}) + (1-\eta_c(\boldsymbol{\hat{\theta}}^{(m)}, \boldsymbol{\hat{\phi}}^{(m)})) \,S_{cu}(t_c \mid \boldsymbol{\hat{\theta}}^{(m)}, \boldsymbol{\hat{\phi}}^{(m)})},\nonumber\\
 \pi(Z=1&\mid \mathcal{D},\boldsymbol{\hat{\theta}}^{(m)}, \boldsymbol{\hat{\phi}}^{(m)}) = 1-\pi(Z=0\mid \mathcal{D},\boldsymbol{\hat{\theta}}^{(m)}).
\end{align*}

\item[(d)] Define $\boldsymbol z^{(m)}=\{ \boldsymbol z_{cen}^{(m)}, \boldsymbol z_{unc} \}$.
\end{itemize}
\end{itemize}


%
\section{Illustrative studies}
\label{sec:5.5}
We considered two benchmark datasets to illustrate  our proposal for estimating mixture cure models via INLA. They are the so-called Eastern Cooperative Oncology Group (ECOG) phase III clinical trial e1684 dataset~\citep{Kirkwood1996} and the bonemarrow transplant study dataset~\citep{Kersey1987}. In both studies, we compared our results with the ones obtained via MCMC methods. Inferences in both studies   were performed on a Windows laptop with an Intel(R) Core(TM) i7-7700 3.60GHz processor. All implementations were made in the \texttt{R} environment (version 3.4.3). We used the \texttt{R-INLA} package for INLA and   \texttt{JAGS} software (version 4.3.0) through the \texttt{rjags} package~\citep{Plummer2003} for MCMC inferences.
\subsection{ECOG study}

\label{sec:5.5.1}
  The ECOG phase III clinical trial was designed to compare    a high dose interferon alpha-2b ($IFN$) regimen against close observation which was the standard therapy  ($ST$) as the postoperative adjuvant treatment~\cite{Kirkwood1996} in high-risk melanoma patients. Data in the analysis included a total   of 284 observations, of which 88 were right-censored. Relapse-free survival (FFS), in years,  was one of   variables of interest in the study and now   our survival variable.  Covariate information included gender, 113 women ($W$) and 171 men ($M$), treatment (144 people in the $IFN$ group and 140 in $ST$), and age ($A$)  (in years and  centered on the sample mean). FFS sample median was 1.24 and 1.36 years in the case of $M$ and $W$, and 1.82 and 0.98   years in the $IFN$ group  and $ST$, respectively.
\subsubsection*{Incidence and latency model.}

  We considered the same CPH mixture cure model stated by the authors  in~\citep{Kirkwood1996}.   The cure proportion for individual $i$th in the \textit{incidence} model was expressed in terms of a binary regression logistic model defined as
\begin{equation}
\label{eqn:5.9}
\mbox{logit}[\eta_{i}(\boldsymbol{\beta}_1 )] \,= \,\beta_{0,1}+\beta_{W,1}\,I_{W}(i)+\beta_{IFN,1}\,I_{IFN}(i)+\beta_{A,1}\,A_i,
\end{equation}
\noindent where $\beta_{0,1}$ represents the reference category, to be a man receiving $ST$ treatment, and  $I_{G}(i)$ is an indicator  variable with value 1 if individual $i$ has the characteristic $G$ and 0 otherwise.

Survival time   for individual $i$  the uncured subpopulation   was modeled by a CPH model with  hazard   function,
\begin{equation}
\label{eqn:5.10}
h_{ui}(t\mid h_{u0}, \boldsymbol{\beta}_2)  =  h_{u0}(t)\,  \mbox{exp}\{\beta_{0,2}+\beta_{W,2}\,I_{W}(i)+\beta_{IFN,2}\,I_{IFN}(i)+\beta_{A,2}\,A_i\} ,
\end{equation}
\noindent with  Weibull  baseline hazard function  $h_{u0}(t)\,=\,\alpha t^{\alpha-1}$.

The model is completed with the elicitation of a prior distribution for all uncertainties it includes. We assume prior independence and select vague normal distributions centered at zero and   variance 1,000  for all the regression coefficients in (\ref{eqn:5.9}) and (\ref{eqn:5.10})  as well as for $\mbox{log}(\lambda)$. The elicited prior distribution for $\alpha$ is the gamma distribution Ga(0.01, 0.01), a very common election in these models
which  baseline hazard function   is specified in terms of a Weibull distribution.

\subsubsection*{Posterior inferences}

Our algorithm configuration included 50 burn-iterations followed by other 450 iterations for inference. In addition, the simulations were thinned by storing one in five draws in order to reduce autocorrelation in the saved sample. The convergence was evaluated by examining whether the estimated conditional (on $z$) marginal log-likelihood   achieved  stability during the iteration steps of the algorithm.

INLA results   were compared to those
obtained via MCMC methods with the JAGS software. A MCMC algorithm was run considering three Markov chains
with 100,000 iterations each and a burn-in period with 20,000 ones.   In
addition, the chains were thinned by storing one in two hundred iterations in order to reduce
autocorrelation in the saved sample and avoid space computer problems. Convergence was assessed
based on the potential scale reduction factor   and the effective number of independent
simulation draws \citep{GelmanRubin1992}.

\begin{table}[h]
\small\sf\centering
\caption{Summary of the INLA and MCMC  approximate
marginal posterior distributions: mean, standard  deviation,  95$\%$ credible interval, and posterior
probability that the subsequent
parameter is positive.
\label{t:table1}}
\begin{tabular}{llcrccc}
\toprule
  & &Parameter&Mean& Sd& 95 $\%$CI & $P( \cdot >0)$\\
\midrule
\textsl{Incidence} &INLA & $\beta_{0,1}$  &-1.200 & 0.235& [-1.676,-0.753]& 0.000 \\
          &     & $\beta_{W,1} $  & 0.061 & 0.275& [-0.483,0.597]& 0.587 \\
          &     & $\beta_{IFN,1} $  &0.573  & 0.271&[0.045,1.107] & 0.983\\
          &     & $\beta_{A,1} $  & -0.015& 0.010&[-0.035,0.005]& 0.076\\
          &MCMC & $\beta_{0,1}$  &-1.220 & 0.239& [-1.701,-0.777]& 0.000 \\
          &     & $\beta_{W,1}$ & 0.058& 0.283& [-0.518,0.595]&0.585\\
          &     & $\beta_{IFN,1}$  &0.573  & 0.271&[0.045,1.107] & 0.983\\
          &     & $\beta_{A,1}$  & -0.015& 0.010&[-0.035,0.005]& 0.076\\
\midrule
\textsl{Latency}&INLA   &  $\alpha$        & 0.918  & 0.052&[0.818,1.022]   & $-$\\
          &            &    exp$\{\beta_{0,2}\}$     & 0.938  & 0.113&[0.729,1.173] &  $-$\\
          &            &$\beta_{W,2}$  &  0.131 & 0.161& [-0.187,0.442] &0.794 \\
          &            &   $\beta_{IFN,2}$  & -0.106 & 0.154&[-0.410,0.195] &0.244 \\
          &            &  $\beta_{A,2}$   & -0.007 & 0.005& [-0.018,0.004]&0.098  \\
          &  MCMC          &    $\alpha$      & 0.909 &0.055&[0.802,1.016]   & $-$ \\
          &            &    exp$\{\beta_{0,2}\}$    &   0.921& 0.114&[0.715,1.152] & $-$\\
          &            &    $\beta_{W,2}$ &  0.133& 0.168& [-0.201,0.437] &0.779 \\
          &            &    $\beta_{IFN,2}$ & -0.108& 0.165&[-0.441,0.209] &0.269 \\
          &            &   $\beta_{A,2}$  & -0.007& 0.006& [-0.018,0.003]&0.102  \\

                                           \bottomrule
\end{tabular}
\end{table}

The number of iterations   needed to accomplish convergence under our proposal
is a fraction than the one in the MCMC configuration. This occurs because our algorithm only
needs to explore the parameter space of the cure indicator variables corresponding to the censored observations, $\mathcal{Z}_{cen}$, and not
the full parameter space  $\mathcal{Z}$. Futhermore, the parameters space of model parameters $(\boldsymbol \theta, \boldsymbol \phi)$ is not explored as their posterior marginals are computed from the conditional posterior marginals obtained with INLA.

Table \ref{t:table1} shows a summary of the INLA and MCMC  approximate  posterior marginal distribution of the parameters
of the mixture cure model estimated.    The  agreement in all the outputs is quite high and confirms that our approach works and provides
similar estimates to MCMC.

The estimation
of the cure proportion as well as the survival profiles for the different  groups of individuals are relevant issues in
the medical context of the study. INLA computes  an approximation to the conditional marginal log-likelihood function $\pi(\mathcal{D}| \boldsymbol z)$ and  
it can be used to select the most likely configuration of the latent vector $\boldsymbol z$ that has
been generated during the sample  process to approximate the posterior distribution of the cure proportion and the survival profiles.  In particular, the
 \texttt{inla.posterior.samples} function in the R-INLA package may  be used to generate  samples from the approximated joint posterior
distribution of the estimated model (we select the most likely model). Additionally,  these samples can subsequently be
  processed to derive approximated posterior distributions for the quantities of interest.

\begin{table}[h]
\small\sf\centering
\caption{Summary of the approximated  INLA   and  MCMC  posterior mean, standard deviation, and    95$\%$ credible interval  of the cure proportion  for  averaged  age individuals   in the four groups of the study. \label{t:table2}}
\begin{tabular}{llccc}
\toprule
  & Group &Mean& Sd& 95 $\%$CI  \\
\midrule
   INLA   &\textit{M-ST}&0.242& 0.042&  [0.166, 0.333]  \\
          &\textit{M-IFN} & 0.363& 0.046& [0.280, 0.453] \\
          &\textit{W-ST} & 0.258& 0.048& [0.172, 0.357] \\
          &\textit{W-IFN} & 0.382& 0.056& [0.278, 0.495]\\
\midrule
 MCMC     & \textit{M-ST} &0.231& 0.042& [0.230, 0.315]   \\
          & \textit{M-IFN} & 0.345& 0.048& [0.252, 0.443]\\
          & \textit{W-ST}  & 0.242 &0.049 &[0.151, 0.346] \\
          & \textit{W-IFN} & 0.358& 0.057& [0.248, 0.475] \\
 \bottomrule
\end{tabular}
\end{table}

Table \ref{t:table2} includes  the INLA and MCMC   posterior mean, standard deviation and 95$\%$ credible interval
of the posterior  distribution of the  cure proportion for  individuals in the four groups of interest: men treated with  the standard therapy \textit{(M-ST)}, men treated with interferon alpha-2b \textit{(M-IFN)}, women in the standard therapy group \textit{(W-ST)}, and $IFN$ women   \textit{(W-IFN)}.  Outcomes from INLA and MCMC   also are in close agreement and highlight that   \textit{(W-IFN)} individuals  present the highest cure proportion
estimates while the lowest values correspond to \textit{M-ST} ones. Differences between treatments   are clinically relevant for both women and men.

\begin{figure}
\centering
\includegraphics[width=4.4cm]{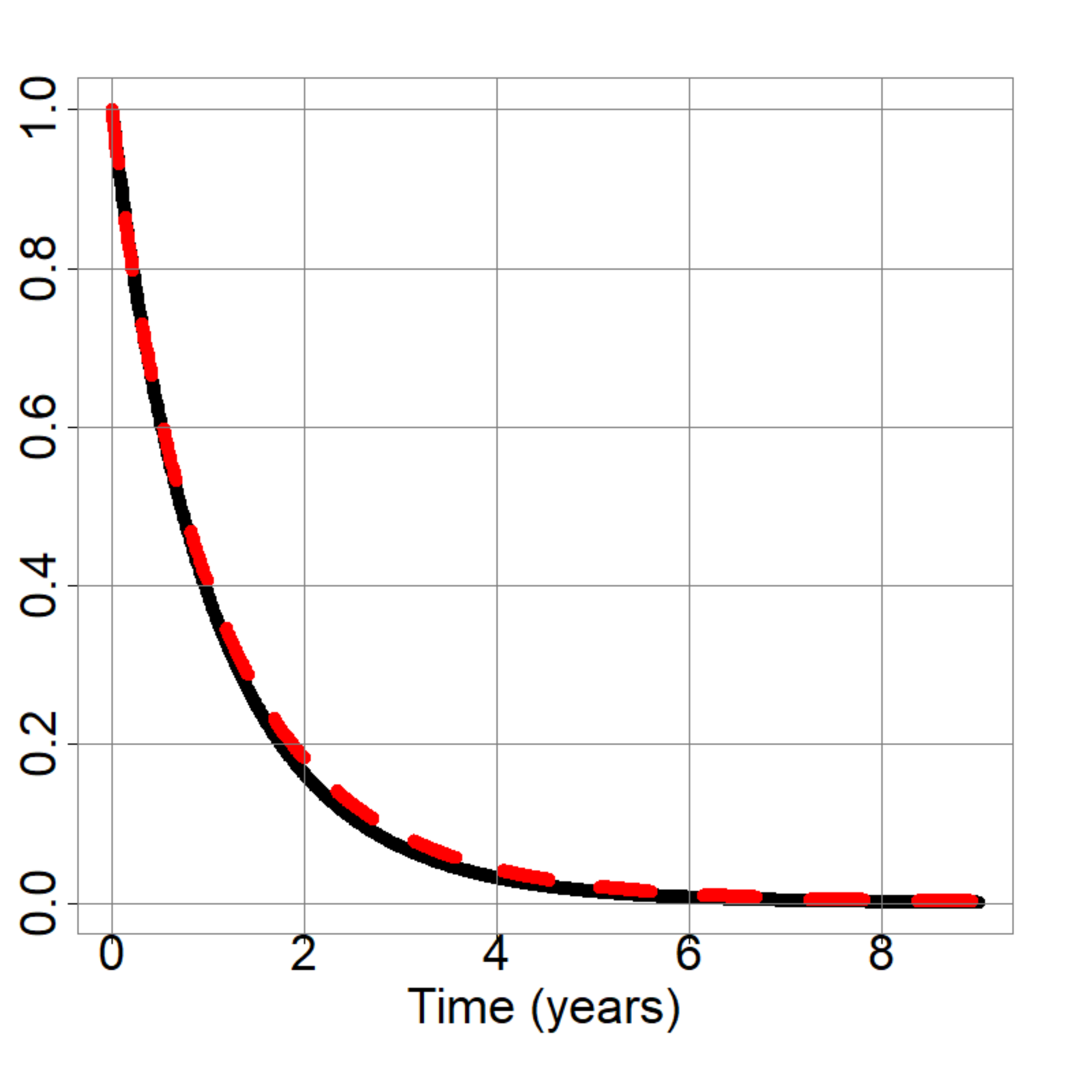}\hspace*{0.4cm}\includegraphics[width=4.4cm]{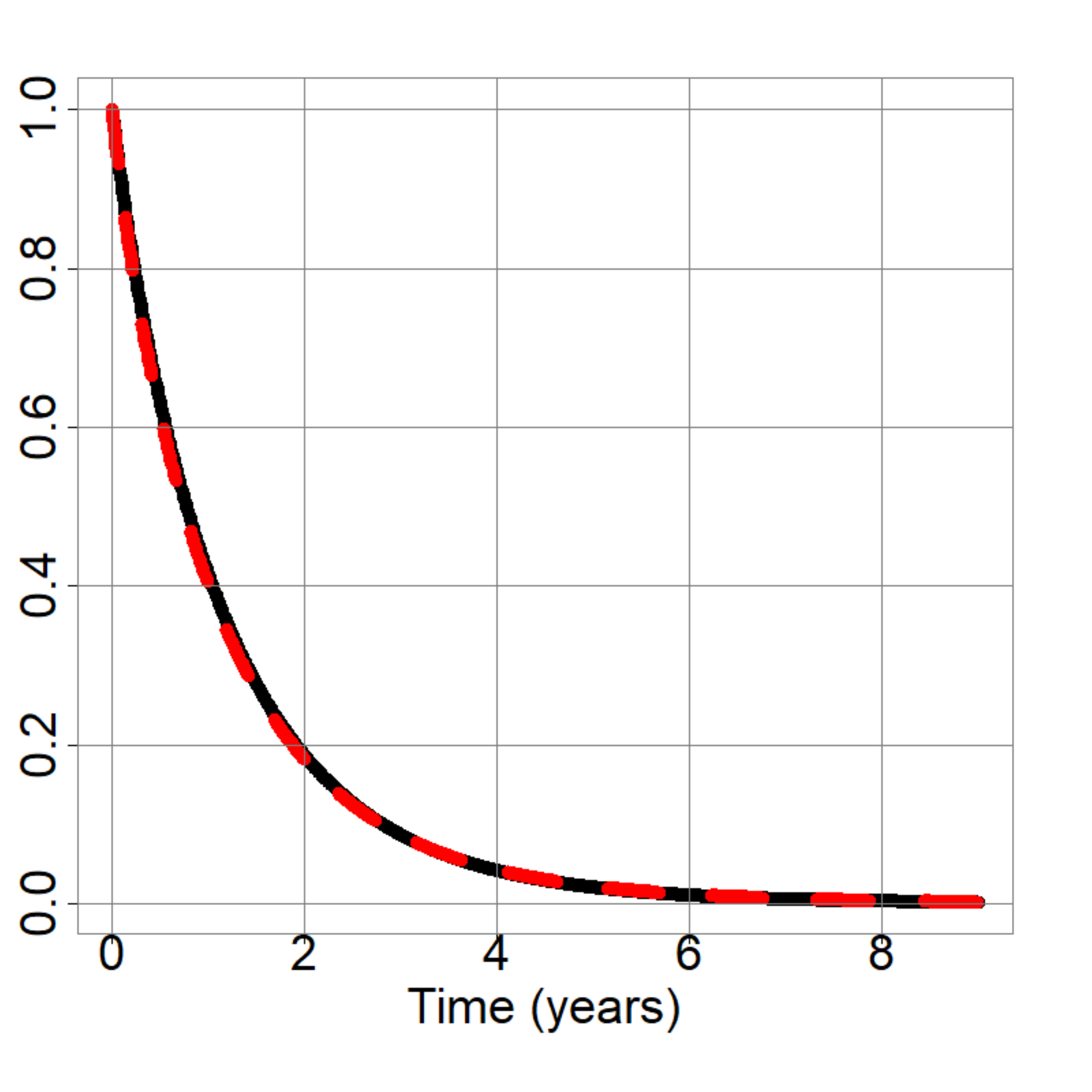}\\
\includegraphics[width=4.4cm]{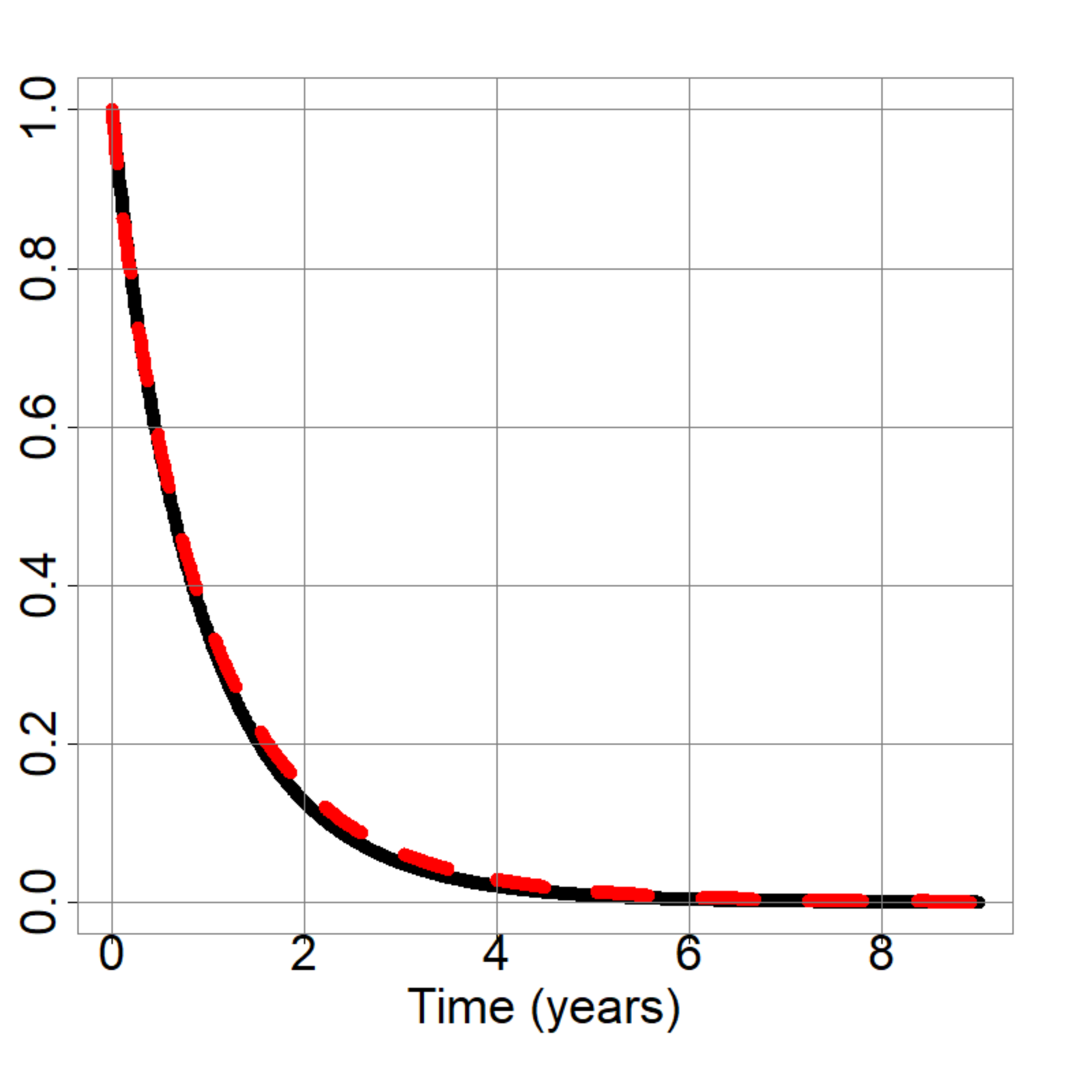}\hspace*{0.4cm}\includegraphics[width=4.4cm]{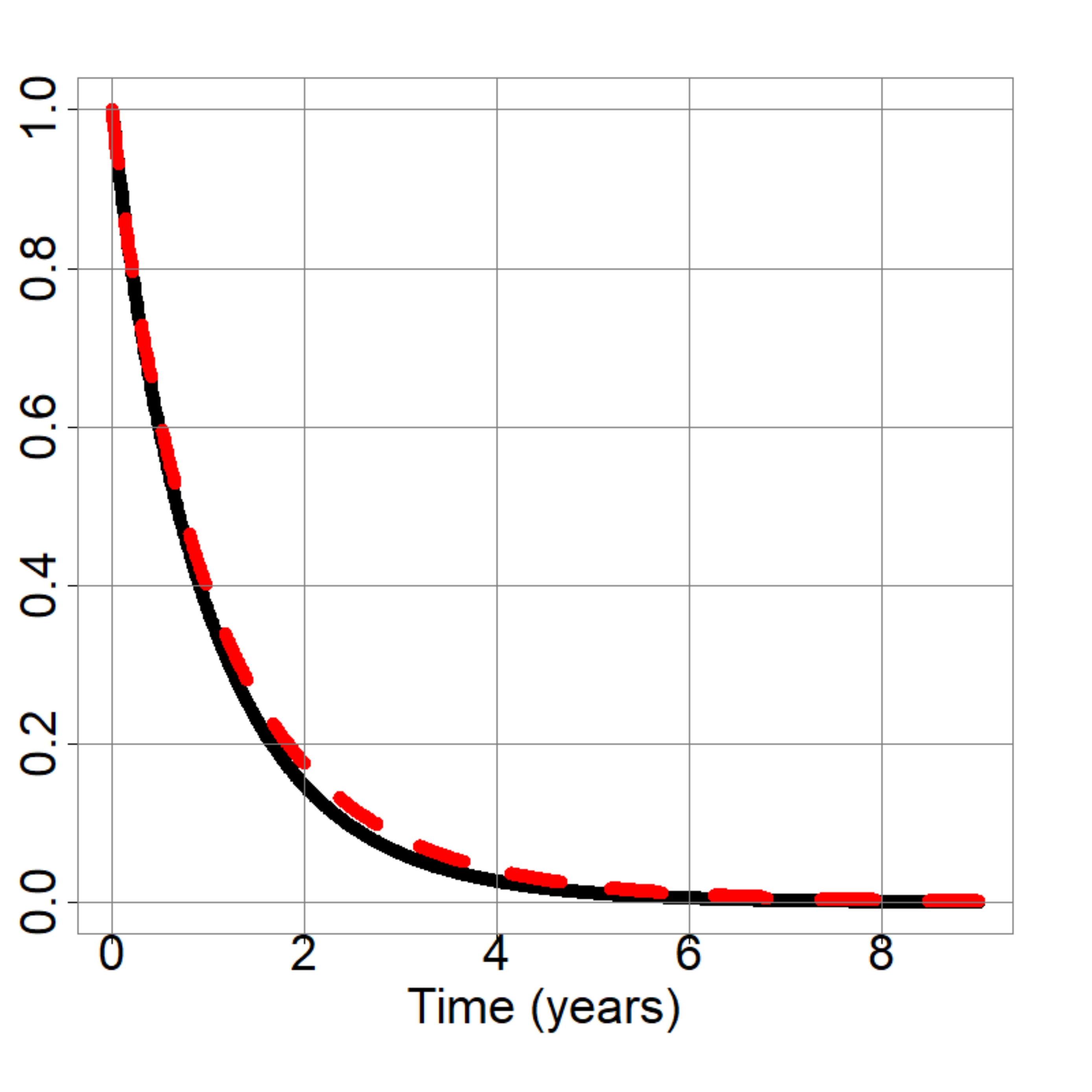}
\caption{Posterior mean of the  uncured  survival function for  individuals in the groups  \textit{M-ST}(on the top left), \textit{M-IFN} (on the top right),  \textit{W-ST} (at the bottom   left),  and \textit{W-IFN} (at the bottom   right)  computed with INLA (black solid line) and MCMC (red dashed line).}
\label{fig:melasurv}
\end{figure}

 Figure~\ref{fig:melasurv} displays  the INLA and MCMC mean of the posterior distribution of the  uncured  survival function for individuals in each of the four groups of interest.  Estimation from both approaches scarcely differs. From a clinical point of view, the survival profiles are very similar among the groups, but it seems that  the best and worst survival expectations  correspond to  \textit{M-IFN} and \textit{W-ST} groups, respectively. We could conclude that the probability of cure is very different among the groups (see Table \ref{t:table2}) but the uncured survival profiles of the individuals in the different groups are very similar (see Figure \ref{fig:melasurv}).

\subsection{Bone marrow transplant study}
\label{sec:5.5.2}
Next, we consider the bone marrow transplant study dataset in~\cite{Kersey1987} to illustrate our proposal for a  Weibull AFTMC model. This study was undertaken to compare autologous and allogeneic marrow transplantation with regard to survival times of patients affected with lymphoblastic leukemia and poor prognosis. A total of $91$ patients were treated with high-doses of chemoradiotherapy and followed-up during a period between 1.4 to 5.0 years. Forty-six patients with a HLA-matched donor received a donor marrow (allogeneic graft)  and 45 patients without a matched donor received their own marrow taken during remission and purged of leukemic cells with the use of monoclonal antibodies (autologous graft). The survival variable of interest was  time to death, in days, which ranged from 11 to 1845 days. Data contain 22 right-censored observations and 69 uncensored. In general, times to death are longer for allogeneic transplant  patients (sample median was 292 days) than for  autologous patients (sample median was 112 days).

  The main goal of the study was to compare both groups, autologous and allogeneic, with regard to the \textit{incidence} and the \textit{latency} models. Covariate  information only contemplates the type of transplant and  was incorporated in both terms of the cure model.

\subsubsection*{Incidence and latency model}
The cure probability for individual $i$th corresponding to the \textit{incidence} model was expressed in terms of a regression logistic model defined as
\begin{equation}
\label{eqn:5.11}
\mbox{logit}[\eta_i(\boldsymbol{\beta} )] \,= \,\beta_{All,1}+\beta_{Aut,1}\,I_{Aut}(i),
\end{equation}
\noindent where $\beta_{All,1}$ represents the effect of the reference category, to be an individual who has received an allogeneic transplant, and  $I_{Aut}(i)$ is an indicator  variable with value 1 whether individual $i$ has had an autologous graft.

Survival time for individual $i$ in the  uncured subpopulation, $T_{ui}$, was modeled by means of a Weibull AFT model    defined   as
\begin{align}
\mbox{log}(T_{ui})&=\beta_{All,2}+\beta_{Aut,2}\,I_{Aut}(i)+\sigma \, \epsilon_i,
\label{eqn:5.12}
\end{align}
\noindent where now $\beta_{All,2}$ represents the effect of  receiving an   allogeneic graft,  and   $\beta_{Aut,1}$   the additional effect for having an autologous transplant.


The model is completed with the elicitation of a prior distribution for all parameters it contains. We assume prior independence and select vague normal distributions centered at zero and   variance 1,000  for all the regression coefficients in the model except for $\alpha=1/\sigma$, for which a Ga(0.01, 0.01) distribution was selected.




\begin{table}[h]
\small\sf\centering
\caption{Summary of the INLA and MCMC  approximate
marginal posterior distributions: mean, standard  deviation,  95$\%$ credible interval, and posterior
probability that the subsequent
parameter is positive.\label{t:table3}}
\begin{tabular}{llcrccc}
\toprule
  & &Parameter&Mean& Sd& 95 $\%$CI & $P( \cdot >0)$\\
\midrule
\textsl{Incidence} &INLA & $\beta_{All,1}$  &-0.988& 0.341& [-1.691,-0.351]& 0.000 \\
          &     & $\beta_{Aut,1} $          & -0.404& 0.505& [-1.407,0.575]&0.211 \\
          &MCMC & $\beta_{All,1}$           &-1.025& 0.355& [-1.763,-0.367]& 0.000   \\
          &     & $\beta_{Aut,1}$           & -0.413& 0.524& [-1.437,0.665]&0.203\\
\midrule
\textsl{Latency} &INLA &   $\beta_{All,2}$  &  -6.372& 0.652& [-7.709,-5.131]&0.000   \\
          &            &   $\beta_{Aut,2}$  &  0.759& 0.262 &[0.247,  1.277]&0.998 \\
          &            &   $\alpha$        &  1.138& 0.103& [0.941,1.343]& $-$ \\
              &MCMC        &   $\beta_{All,2}$  & -6.305& 0.631& [-7.572,-5.118]&0.000  \\
          &            &   $\beta_{Aut,2}$  &  0.754& 0.267 &[0.238,  1.287]&1.000 \\
          &            &   $\alpha$        &  1.124& 0.101& [0.934,1.325]& $-$ \\    \bottomrule
\end{tabular}
\end{table}

\subsubsection*{Posterior inferences}

Our algorithm configuration for this  model included 20 burn-in iterations and other 180  for inference. In addition, the simulations were thinned by storing every 2nd draws in order to reduce autocorrelation in the saved sample. Convergence was evaluated by examining whether the conditional   marginal log-likelihood estimates achieved  stability during the iteration steps of our algorithm.

\begin{table}[h]
\small\sf\centering
\caption{Summary of the approximated  INLA   and  MCMC  posterior mean, standard deviation, and    95$\%$ credible interval  of the cure proportion  for  allogeneic and autologous graft patients. \label{t:table4}}
\begin{tabular}{llccc}
\toprule
  & Group &Mean& Sd& 95 $\%$CI  \\
\midrule
   INLA   &\textit{All}    &0.286& 0.067& [0.170,0.428]  \\
          &\textit{Aut}  & 0.205& 0.060& [0.107,0.342]\\
\midrule
 MCMC     & \textit{All}  &0.270& 0.067& [0.146,0.410]  \\
          & \textit{Aut}  & 0.198& 0.057& [0.094,0.319]   \\
 \bottomrule
\end{tabular}
\end{table}

MCMC simulation was run considering three Markov chains with 200,000 iterations and a burn-in period with 40,000 iterations. The chains were thinned by storing every 400th iteration to reduce autocorrelation in the saved sample and avoid space computer problems. Convergence was also here assessed via the potential scale reduction factor  and the effective number of independent simulation draw~\citep{GelmanRubin1992}. As   in the ECOG study,     our proposed method here also needed less iterations than MCMC configuration to reach convergence and accurate results.

Table~\ref{t:table3} shows the INLA and MCMC   mean, standard deviation and 95$\%$
credible interval of the posterior distribution of the cure proportion for allogeneic and autologous transplant patients. INLA and MCMC results are very similar.

\begin{figure}
\centering
\includegraphics[width=4.4cm]{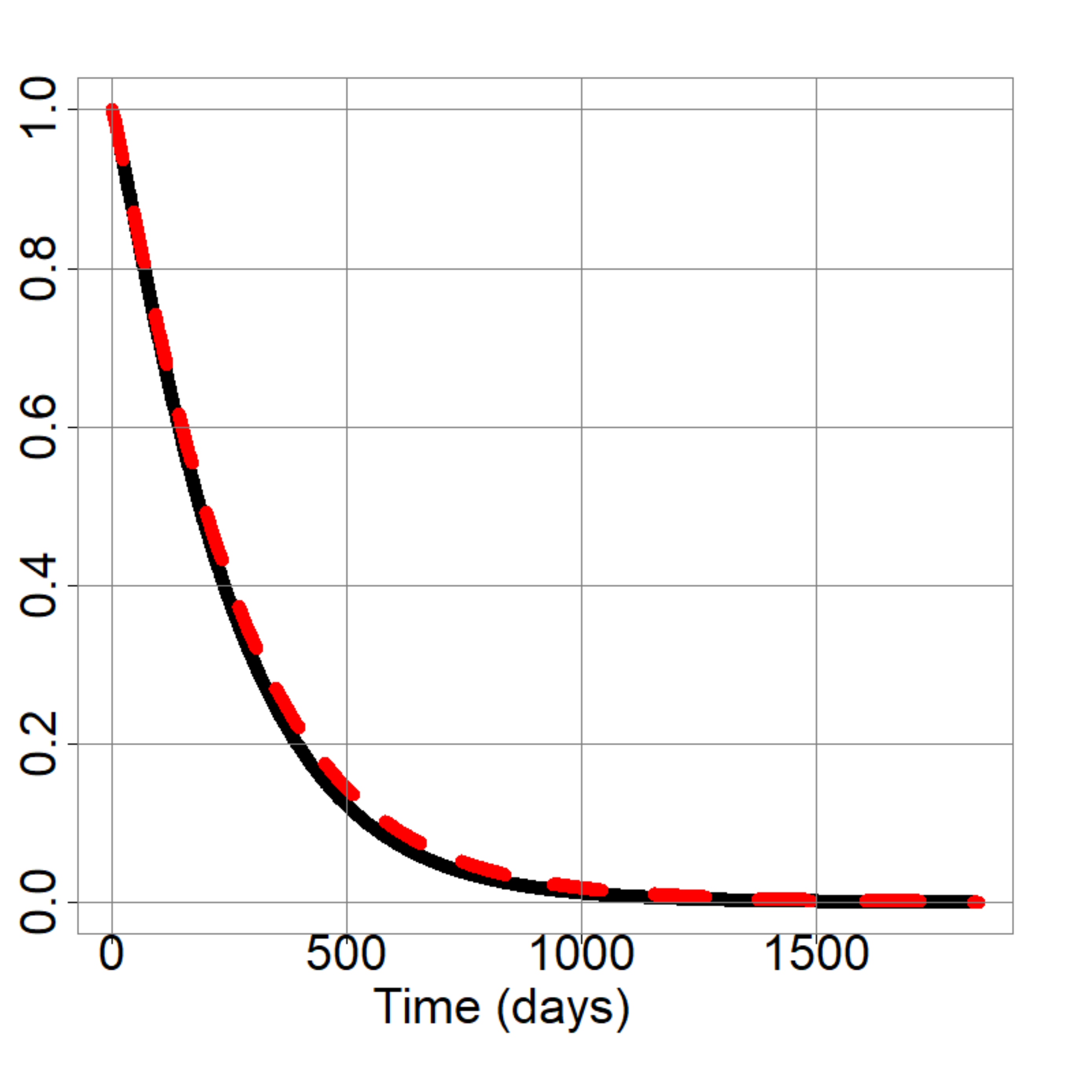}\hspace*{0.4cm}\includegraphics[width=4.4cm]{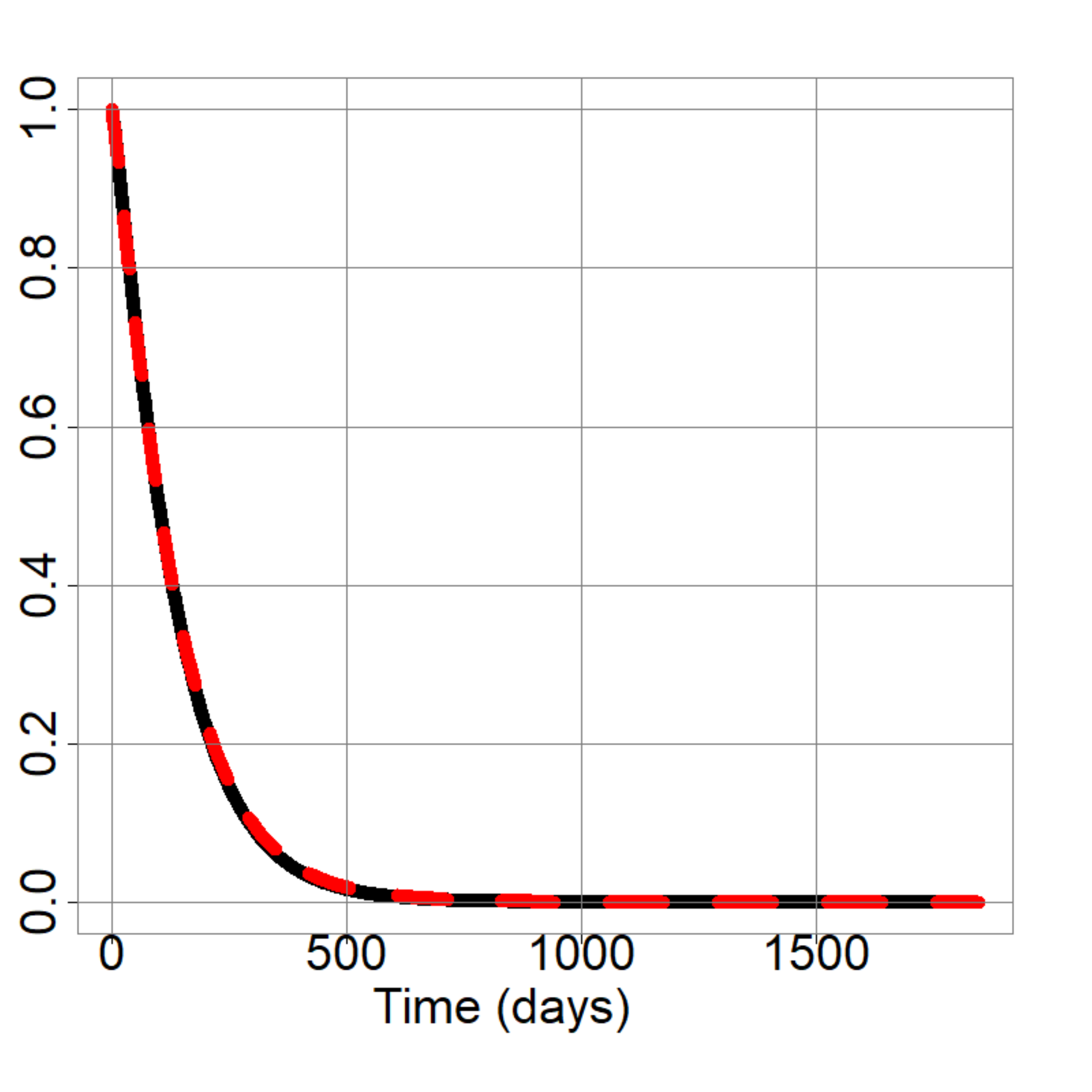}\\
\caption{Posterior mean of the uncured survival function for \textit{Allogeneic} (on the left) and \textit{Autologous} (on the right) transplanted patients computed from INLA (black solid line) and MCMC (red dashed line)}
\label{fig:bonesurv}
\end{figure}

In the case of the estimation of derived quantities of interest, we proceed analogously to the ECOG study. We estimate the INLA and MCMC posterior distribution for the cure proportion for allogeneic and autologous transplant patients (Table~\ref{t:table4}) as well as the subsequent posterior mean of the uncured survival function (Figure~\ref{fig:bonesurv}). Outcomes also now present scarce differences and underline that allogeneic  transplanted patients seem to have cure proportion levels higher than the  ones for autologous patients, although we also appreciate   a very broad degree of overlap.

\section{Conclusions}
\label{sec:5.6}


 This paper discussed an INLA approach  for dealing with  mixture cure models based on a general
procedure by \cite{Bivand2014, GomezRubio2017, GomezRubio2018} that extends INLA to   finite mixture models. We introduced    latent indicators in the inferential process for classifying   individuals   in the cured and uncured subpopulations,   and   approximated the relevant posterior distribution via  Gibbs sampling.  In particular, we use modal Gibbs sampling \citep{GomezRubio2018} and    INLA to fit the  marginal posterior distribution of each relevant element given the latent indicator variable that identifies each individual in the cure or uncured population.

Two specific benchmark datasets  from the field of medicine have been considered to illustrate our proposal. In both cases, the results support its viability and good performance, and almost entirely agree with the  MCMC results. Remarkably, our proposal also shows other interesting properties  such as the lower number of iterations to reach convergence and the convenient exploration of the parametric space of the latent indicators. Furthermore, the use of INLA to fit conditional models does not force the use of conjugate priors in the Gibbs sampler and avoids  label
switchings problems usually caused  by symmetry in the likelihood function of the model parameters~\citep{Stephens2000}.

On the other hand, MCMC   provides, at the moment, slightly faster computational times and consequently, more research would be necessary to minimize computational efforts and storage requirements. Note that INLA estimates two complete  models, incidence  and  latency,  in each iteration. This leads to an important  computational burden because two complete processes in each iteration were generated thus producing
  new temporary files and other secondary elements. So, if we limit the default outcomes provided by INLA and we define prior distributions based on the inferences from the previous iteration, computational savings could be achieved.

\vspace{1cm}

\noindent
{\bf Declaration of Conflicting Interests:} The authors declare that there is no conflict of interest.

\begin{acks}
L\'azaro's research was supported by a predoctoral FPU fellowship
(FPU2013/02042) from the Spanish Ministry of Education, Culture and Sports. This paper was partially   funded by grant MTM2016-77501-P from the Spanish Ministry of
Economy and Competitiveness co-financed with FEDER funds, and 
a grant funded by Consejer\'ia de
Educaci\'on, Cultura y Deportes (JCCM, Spain) and FEDER.
\end{acks}


\bibliographystyle{SageH}
\bibliography{survival}

\end{document}